\documentclass{class_HNP07}
\def\Slash#1{/\hspace{-0.23cm}{#1}} 
\newcommand{\beq}{\begin{eqnarray}}
\newcommand{\eeq}{\end{eqnarray}}
\newcommand{\bra}{\langle}
\newcommand{\del}{\partial}

\def\beq{\begin{eqnarray}}
\def\eeq{\end{eqnarray}}
\def\nn{\nonumber}

\def\ln{\mbox{ln}}

\def\befc{\begin{figure}[h]\begin{center}}
\def\eefc{\end{center}\end{figure}}

\def\bra{\langle}    \def\ket{\rangle}

\def\feynint2{\int_0^12xdx\int_0^1dy}

\def\ms{M_s}
\def\ma{M_A}
\def\mq{m_q}

\def\bra{\langle}    \def\ket{\rangle}
\def\del{\partial}
\def\ms{M_S} \def\mq{m_q} \def\ma{M_A}

\def\ds{\displaystyle}

\def\diag{\mbox{diag}}
\begin{document}
\setcounter{page}{1}
\title{ Structure of the Roper Resonance with Diquark Correlations} 
\author{Keitaro Nagata$^{1,2}$, Atsushi Hosaka$^2$}
\address{
$^1$Chung Yuan Christian University, Department of Physics,\\
Chung-Li 320, Taiwan \\
$^2$Research Center for Nuclear Physics,  Osaka University, \\
 Ibaraki 567-0047, Japan.\\
nagata@phys.cycu.edu.tw, nagata@rcnp.osaka-u.ac.jp}
\maketitle
\abstracts{
We study electromagnetic properties of the nucleon and Roper
resonance in a chiral quark-diquark model including two kinds of 
diquarks needed to describe the nucleon: scalar and axial-vector diquarks. 
The nucleon and Roper resonance 
are described as superpositions of two quark-diquark bound 
states of a quark and a scalar diquark and of a quark and an 
axial-vector diquark. Electromagnetic form factors of the nucleon 
and Roper resonance are obtained from one-loop diagrams where 
the quark and diquarks are coupled by a photon.
We include the effects of intrinsic properties of the diquarks: 
the intrinsic form factors both of the diquarks and the anomalous magnetic 
moment of the axial-vector diquark. 
The electric form factors of the proton and neutron reasonably agree with 
the experiments due to the inclusions of the diquark sizes, while the magnetic moments
become smaller than the experimental values because of the scalar dominance
in the nucleon.  The charge radii of the Roper resonance are almost 
comparable with those of the nucleon.
}
\section{Introduction}

Diquarks or diquark correlations, which are strongly attracted two quarks, 
are known to play important roles in several phenomena, such as the ratio of 
the structure functions $\ds{\lim_{x\to 1}} F_2^n(x)/F_2^p(x)\to 1/4$,  $\Delta I=1/2$ 
rule in semi-leptonic weak decays, recent exotic studies and so 
on~\cite{Selem:2006nd}. 
The existence and applications of the diquark correlations are of recent interests
and have been investigated extensively. 
Recently, we have proposed a quark-diquark description of the Roper resonance 
as a partner of the nucleon~\cite{Nagata:2005qb}, where diquarks played a crucial 
role on the existence of the Roper resonance and its excitation energy.

Diquark models describe the nucleon as a superposition of the two kinds of the 
quark-diquark bound states: a bound state of a scalar diquark and  quark and 
of an axial-vector diquark and quark. If the spin-flavor symmetry $SU(4)_{SF}$ is an exact symmetry, the 
nucleon is restricted to be the linear combination of those quark-diquark bound 
states with the equal weight due to the Pauli-principle. 
On the other hand, if $SU(4)_{SF}$ is not an exact one, the nucleon 
can be an arbitrary linear combination. In addition, not only the nucleon but
also the other state orthogonal to the nucleon are allowed to appear as a physical state.
Most famous effect of the spin-flavor violation is the color magnetic 
interaction between quarks, as it generates the mass difference between the 
nucleon and $\Delta(1232)$ being about 300 MeV.
Hence the mass difference  between the nucleon and the other 
state is expected to be the typical order of $SU(4)_{SF}$ violation $M_\Delta-M_N\sim 300$, 
which is comparably smaller than the nodal excitation energy of the Roper resonance $2\hbar \omega$.

In Ref.~\cite{Nagata:2005qb}, we showed that the two 
quark-diquark bound states were identified with the nucleon and Roper resonance
by evaluating their masses in the framework of a chiral quark-diquark model.
The chiral quark-diquark model incorporates the two kinds of diquarks. 
We assumed the mass difference between 
the scalar and axial-vector diquarks to be an order of the color magnetic interactions.
Hence we have this amount of the mass difference between the pure bound 
states of a quark and a scalar diquark and of a quark and an axial-vector 
diquark. Introducing a mixing interaction for those two channels
causes an additional mass difference, as we are familiar with two-level 
problems in quantum mechanics.
In this model setup, we showed that the mass difference between the 
two states were about 300-600 MeV.
Assigning the low-lying state to the nucleon, the other state with 
the excitation energy about 500 MeV is naturally identified with 
the Roper resonance. 

In the quark-diquark description, the difference of the diquark correlations
between the scalar and axial-vector types is a root cause of
the violation of the $SU(4)_{SF}$ symmetry and of the appearance of the Roper resonance.
In this sense the Roper resonance is a spin-partner of the nucleon  
having different diquark components of different spins, which is much 
different from conventional descriptions, the Nambu-Goldstone boson exchange~\cite{Glozman:1995fu}, 
collective excitations such as the vibrational~\cite{Brown:1983ib}, 
and rotational mode~\cite{Hosaka:1999aa}.
If the Roper is described by diquark correlations, its nature would be
much different.

In the present paper, we extend the chiral quark-diquark model to the 
electromagnetic properties of the nucleon and Roper resonance.
The method of the chiral quark-diquark model resembles the Faddeev approach of the NJL model 
with the static approximation, where Mineo et al.~\cite{Mineo:2002bg} investigated the electromagnetic properties 
of the nucleon. 
Hence we expect that the chiral quark-diquark model describes the electromagnetic 
properties of the nucleon.
Since the Roper resonance appears with the nucleon, the properties of the 
Roper resonance is obtained simultaneously with the nucleon.

This paper is organized as follows. In section~\ref{sec:framework} we 
construct the quark-diquark model for the quark and diquarks with the photon  
introduced via the covariant derivatives. 
In terms of the path-integral auxiliary field method, which
we call the path-integral hadronization, the quark and diquark fields
are transformed into auxiliary baryon fields.
Expanding a Tr ln term resulted from the path-integral hadronization, 
we obtain various loop diagrams describing hadron properties.
We briefly review the kinetic and mass terms studied in 
Ref.~\cite{Nagata:2005qb}, where the auxiliary baryon fields are identified 
with the nucleon and Roper resonance. The electromagnetic properties of 
the nucleon and Roper resonance 
are obtained from loop diagrams where the quark and diquarks are coupled with a photon.
We show numerical results in section~\ref{sec:em}.  
The final section is devoted to a summary.
\section{Framework}\label{sec:framework}

We construct the chiral quark-diquark model and derive thek kinetic and mass terms, and 
electromagnetic properties of the nucleon and Roper resonance.

\subsection{Chiral Quark-Diquark Model}
\label{sec:model}
We start from the SU(2)$_R\times$ SU(2)$_L$ chiral quark-diquark model 
of~\cite{AbuRaddad:2002pw,Nagata:2003gg,Nagata:2004ky}, 
\begin{eqnarray}
{\cal L}_{qD} &=& \bar{\chi}_c(i\rlap/\del - m_q) \chi_c +\;D^\dag_c (\del^2 + M_S^2)D_c
\nonumber\\
&& +
{\vec{D}^{\dag\;\mu}}_c 
\left[  (\del^2 + M_A^2)g_{\mu \nu} - \del_\mu \del_\nu\right]
\vec{D}^{\nu}_c +{\cal L}_{int},
\label{lsemibos}
\end{eqnarray}
where $\chi_c$, $D_c$ and $\vec{D}_{\mu c}$ are the constituent quark, scalar
diquark and axial-vector diquark fields,  $m_q$, $M_S$ and $M_A$
are their masses. The indices $c$ represent the color. Note that the 
diquarks microscopically correspond to the quark bi-linears: $D_c\sim
\epsilon_{abc}\tilde{\chi}_b\chi_c,\ \vec{D}_{\mu c}\sim \epsilon_{abc}\tilde{\chi}_b 
\gamma_\mu \gamma_5 \vec{\tau}\chi_c$, where 
$\tilde{\chi}=\chi^T C\gamma_5 i\tau_2$.  
Both the diquarks belong to 
color anti-triplets and baryons to singlets~\cite{AbuRaddad:2002pw,Nagata:2004ky,Nagata:2007di}.
The term ${\cal L}_{int}$ is the quark-diquark interaction, which is written as
\begin{eqnarray}
{\cal L}_{int}&=&G_S\bar{\chi}_cD^\dagger_c
D_{c^\prime}\chi_{c^\prime}+v(\bar{\chi}_cD^\dagger_c\gamma^\mu\gamma^5
\vec{\tau}\cdot\vec{D}_{\mu c^\prime} \chi_{c^\prime}+\bar{\chi}_c\gamma^\mu\gamma^5
\vec{\tau}\cdot\vec{D}^\dagger_{\mu c}
D_{c^\prime}\chi_{c^\prime})\nonumber \\
&&+G_A\bar{\chi}_c\gamma^\mu\gamma^5
\vec{\tau}\cdot\vec{D}^\dagger_{\mu c}
\vec{\tau}\cdot\vec{D}_{\nu c^\prime}\gamma^\nu\gamma^5 \chi_{c^\prime},
\label{eq:twoc}
\end{eqnarray}
where $G_S$ and $G_A$ are the coupling constants for the quark and scalar
diquark, and for the quark and axial-vector diquark, while  
$v$ causes the mixing between the scalar and axial-vector channels in the nucleon. 
The term ${\cal L}_{int}$ is invariant under the non-linear chiral 
transformation~\cite{AbuRaddad:2002pw,Nagata:2004ky}.

The quark-diquark interaction terms are gathered in a simple form
\begin{eqnarray}
{\cal L}_{int}=\bar{\psi}\hat{G}\psi,
\end{eqnarray}
where we employ matrix notations,
\begin{eqnarray}
\psi&=&\left(\begin{array}{c} D\chi\\ 
\vec{D}_\mu\cdot\vec{\tau}\gamma^\mu\gamma^5\chi\end{array}\right),
\bar{\psi}=
\left(\begin{array}{cc} 
\bar{\chi}D^\dagger, & 
\bar{\chi}\vec{D}_\mu^\dagger\cdot\vec{\tau} \gamma^\mu \gamma^5 \end{array} \right),
\label{eq:matrixfield}
\hat{G}=\left(\begin{array}{cc} G_S & v \\ v & G_A \end{array}\right).
\end{eqnarray}
Note that  the quark-diquark fields $D\chi$ and 
$\vec{D}_\mu\cdot \vec{\tau}\gamma^\mu\gamma^5 \chi$ have the quantum number of 
the nucleon. 
Here and from now on, we omit the color indices.
 
The photon field $A_\mu$ is introduced via the gauge couplings
$\hat{D}_\mu=\partial_\mu+i Q A_\mu$
\begin{eqnarray}
{\cal L}_{qD} = \bar{\chi} \left(S\right)^{-1} \chi \;+\;
D^\dag \left(\Delta\right)^{-1} D
 +
D^{\dag\;\alpha i}
\left( \Delta_{\alpha\beta}^{ij}\right)^{-1}
D^{\beta j} +\bar{\psi}\hat{G}\psi,
\label{eq:lagem}
\end{eqnarray}
where we have introduced the shorthand notations $S,\; \Delta$ and 
$\Delta_{\mu\nu}$:
\begin{eqnarray}
\left(S\right)^{-1}&=&\left(S^0\right)^{-1}+\Gamma_{q}^\mu A_\mu ,\\
\left(\Delta\right)^{-1}&=&\left(\Delta^0\right)^{-1}+\Gamma_{S}^\mu A_\mu ,\\
\left(\Delta_{\alpha\beta}^{ij}\right)^{-1}&=&\left(\delta^{ij} \Delta_{\alpha\beta}^{0}\right)^{-1}+\Gamma_{A\alpha\beta}^{\mu ij} A_\mu.
\label{eq:propax}
\end{eqnarray}
$S^0$, $\Delta^0$ and $\Delta_{\alpha\beta}^0$ are the bare 
propagators of the quark, scalar diquark and axial-vector diquark
and $\Gamma_q^\mu$, $\Gamma_S^\mu$ and $\Gamma_{A\alpha\beta}^{\mu ij}$ 
are the electromagnetic vertices for the quark, scalar and 
axial-vector diquarks given as
\begin{eqnarray}
\Gamma_{q}^\mu&=&-Q_q\gamma^\mu,\\
\Gamma_{S}^\mu(p_1,\; p_2)&=&-Q_S(p_1+p_2)^\mu,\\
\Gamma_{A\alpha\beta}^{\mu ij}(p_1,\; p_2)&=&- Q_A^{ij}\left[g_\alpha^\mu p_{2\beta}+g_\beta^\mu p_{1\alpha}-g_{\alpha\beta}(p_1+p_2)^\mu\right.\nonumber\\
&&+\left.\kappa(g_\alpha^\mu(p_2-p_1)_\beta
+g_\beta^\mu(p_1-p_2)_\alpha)\right].
\label{eq:anmls}
\end{eqnarray}
Here $p_1$ and $p_2$ are the momenta of the incoming and outgoing diquarks, and
$\kappa$ is an anomalous magnetic moment of the axial-vector diquark
~\cite{Kroll:1993zx,He:2005hw}.
$Q_q,\;Q_S$ and $Q_A^{ij}$ are the charges of the quark, scalar diquark and 
axial-vector diquark, as defined by $Q_q=\tau_0/6+\tau_3/2$,
$Q_S=1/3,\;Q_A^{ij}=\diag(4/3,-2/3,1/3)$, where  $ D^{\mu i}=(D^-,\;D^+,\;D^0)$.

Next we carry out the path-integral hadronization  to introduce baryon 
fields.
To begin with, we introduce auxiliary baryon fields in Eq.~(\ref{eq:lagem}),
\begin{eqnarray}
{\cal L}_{qDB}= {\cal L}_{qD}-\bar{B}\hat{G}^{-1}B.
\label{eq:lag_hadronized}
\end{eqnarray}
$B=(B_1, B_2)^T$ is a two component auxiliary baryon field, which 
correspond to $B_1\sim D\chi$ 
and $B_2\sim \vec{\tau}\cdot\vec{D}_\mu\gamma^\mu\gamma^5\chi$.
Eliminating the quark and diquark fields using  the path-integral identities~\cite{AbuRaddad:2002pw}, 
 Eq.~(\ref{eq:lag_hadronized}) is reduced to
\begin{equation}
{\cal L}_ B=-\bar{B}\hat{G}^{-1}B+i\mbox{Tr}\; \ln(1-A).
\label{eq:trln}
\end{equation}
Here the matrix $A$ is defined by
\begin{eqnarray}
A&=&\left(\begin{array}{cc} a_{11} & a_{12}\\ a_{21}& a_{22}\end{array}\right),\\
a_{11}&=&\Delta^T \bar{B}_1 S B_1,\\
a_{12}&=&\Delta^T\bar{B}_2 \tau^i \gamma^\nu\gamma^5  S B_1,\\
a_{21}&=&(\Delta_{\rho\nu}^{lj})^T \bar{B}_1 S \gamma^\mu\gamma^5\tau^j  B_2,\\
a_{22}&=&(\Delta_{\rho\nu}^{lj})^T \bar{B}_2\gamma^\nu\gamma^5\tau^i S \gamma^\mu\gamma^5\tau^j B_2.
\end{eqnarray}%
Though the Lagrangian Eq.~(\ref{eq:trln}) looks simple, it 
contains many physical ingredients, which are explicitly obtained by 
the expansion,
\begin{eqnarray}
\mbox{Tr}\;\ln (1-A)=-\left(A+\frac12 A^2+\frac13 A^3+\cdots\right).
\label{eq:trlgex}
\end{eqnarray}
and by the expansion of the propagators Eqs.~(\ref{eq:propax}), for instance for the
quark,  
\begin{eqnarray}
S=S_0-S_0(\Gamma_{q}^\mu A_\mu)S_0+\cdots.
\label{eq:qex}
\end{eqnarray}
Similarly, the propagators of the scalar and axial-vector diquarks are 
expanded in a series of the photon fields.
Since Eq.~(\ref{eq:trlgex}) is the expansion in powers of the baryon 
fields,  the first term describes static properties of the baryons, 
the second term baryon-baryon interaction. 
While Eq.~(\ref{eq:qex}) is in powers of the photon fields.
In the subsequent sections, we study mass terms 
and electromagnetic couplings of the baryons by taking relevant terms of 
these expansions.

\subsection{Mass Terms}
Due to the nature of the auxiliary fields, 
$B_{1,2}$ do not have the kinetic and mass terms intrinsically. 
They are obtained from loop diagrams comprised of the 
quark and diquarks, then $B_{1,2}$ are identified with the 
physical states. 

\begin{figure}[tbh]                                                       
\centering{                                                               
\includegraphics[width=4cm]{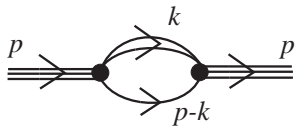}
\begin{minipage}{11cm}                                                    
   \caption{\small A diagrammatic representation of the baryon 
   self-energies. The single, double and triple lines represent the quark,  
   diquark and baryon respectively. The blobs represent the three point  
   quark-diquark-baryon interactions.}                                    
   \label{fig:self}                                                       
 \end{minipage}}                                                          
\end{figure}

Taking the first terms both of the expansions Eqs.~(\ref{eq:trlgex}) and 
(\ref{eq:qex}), we obtain one-loop diagrams shown in Fig.~(\ref{fig:self}). 
Up to this order, the Lagrangian Eq.~(\ref{eq:trln}) is written as
\begin{eqnarray}
{\cal L}=
\bar{B}\left(\begin{array}{cc} 
\Sigma_S(p) & 0 \\ 0 & \Sigma_A(p)\end{array}\right)B 
- \frac{1}{|\hat{G}|}\bar{B}\left(\begin{array}{cc} G_A 
& -v \\ -v & G_S \end{array}\right) B,
\label{eq:lag2}
\end{eqnarray}
where $|\hat{G}|=\mbox{det} \hat{G}=G_SG_A-v^2$.
The scalar and axial-vector diquark contributions to the
self-energies, $\Sigma_S$ and $\Sigma_A$, are given as
\begin{eqnarray}
\Sigma_S(p)&=&-i N_c\int\frac{d^4k}{(2\pi)^4}
\frac{1}{k^2-\ms^2}\frac{\Slash{p}-\Slash{k}+\mq}{(p-k)^2-\mq^2},
\label{eqn:selfs}\\
\Sigma_A(p)&=&-iN_c\int\frac{d^4k}{(2\pi)^4}
\frac{k^\mu k^\nu/\ma^2-g^{\mu\nu}}{k^2-\ma^2}\gamma_\nu\gamma_5\tau_i
\frac{\Slash{p}-\Slash{k}+\mq}{(p-k)^2-\mq^2}\tau_i\gamma_\mu\gamma_5 . 
\label{eqn:selfa}%
\end{eqnarray}%
Here $N_c$ is the number of colors. 
We regularize these self-energies by the three momentum cutoff scheme following the previous 
work~\cite{Nagata:2004ky}.
%
%
With the three momentum cutoff,  $\Sigma_S$ and $\Sigma_A$ are
decomposed into Lorentz scalar and vector $(\gamma^0)$ parts as
\begin{eqnarray}
\Sigma_S(p_0)-\frac{1}{|\hat{G}|}G_A&=&Z_S^{-1}(p_0\gamma^0-a_S),
\label{SigmaS}\\
\Sigma_A(p_0)-\frac{1}{|\hat{G}|}G_S&=&Z_A^{-1}(p_0\gamma^0-a_A),
\label{SigmaA}
\end{eqnarray}\label{Sigma}%
where we employ the baryon rest frame $p_\mu=(p_0,\ \vec{0})$.
The bare baryon fields $B_{1,2}$ are now renormalized as
\begin{eqnarray}
\left(\begin{array}{c} B_1 \\ B_2 \end{array}\right)
=\left(\begin{array}{c}\sqrt{Z_S} B_1^\prime\\
\sqrt{Z_A} B_2^\prime\end{array}\right),
\end{eqnarray}
with which the Lagrangian (\ref{eq:lag2}) is reduced to
\begin{equation}
{\cal L}=\bar{B}^\prime(p_0\gamma^0 - \hat M )B^\prime\, .
\label{eq:kinBprime}
\end{equation}
Here the mass matrix $\hat{M}$ is given as
\begin{eqnarray}
\hat{M}=\left(\begin{array}{cc} a_S & -\sqrt{Z_S Z_A}
\frac{v}{|\hat{G}|}\\ -\sqrt{Z_S Z_A} \frac{v}{|\hat{G}|} & a_A
\end{array}\right).
\label{eq:mmatrix}
\end{eqnarray}
When there is no mixing interaction $(v=0)$, $B_{1,2}^\prime$
become the physical states
with $a_{S}$ and $a_{A}$ being their masses.
In the presence of the mixing, the physical states are obtained
by the diagonalization of the mass matrix with an unitary transformation:
\begin{eqnarray}
B^\prime =U^\dagger N, \;  
U\hat{M}U^\dagger = \diag(M_1,\;M_2).
\label{eq:unitary}
\end{eqnarray}
Finally, Eq.~(\ref{eq:kinBprime}) becomes
\begin{equation}
{\cal L}=\bar{N_1}(p_0\gamma^0-M_1)N_1+\bar{N_2}(p_0\gamma^0-M_2)N_2,
\end{equation}
where the eigenvalues $M_{1,2}$ and eigenvectors 
$N=(N_1,\ N_2)^T$ are obtained as
\begin{eqnarray}
M_{1,2}&=&\frac12\left[a_S+a_A\pm
\sqrt{(a_S-a_A)^2+4Z_S Z_A
\left(\frac{v}{|\hat{G}|}\right)^2}\;\right]\, 
\label{eq:det_mass}, \\
N_1&=&\cos\phi B_1^\prime+\sin\phi B_2^\prime\\
N_2&=&-\sin\phi B_1^\prime+\cos\phi B_2^\prime \, , 
\end{eqnarray}
and the mixing angle $\phi$ is given by 
\begin{equation}
\tan\phi=\frac{|\hat{G}|\sqrt{Z_S Z_A}(a_S-M_1)}{v} \, .
\end{equation}
Eq. (\ref{eq:det_mass}) should be read as a self-consistent 
equation where the quantities on the right hand side 
are functions of $M_{1,2}$.  
The equations are then equivalent to the Schr\"odinger 
equation for the quark-diquark system interacting through 
the delta function type interaction with a suitable cutoff.

\subsection{Electromagnetic couplings of the nucleon and Roper resonance}
\begin{figure}[tbh]
\centering{
\includegraphics[width=4cm]{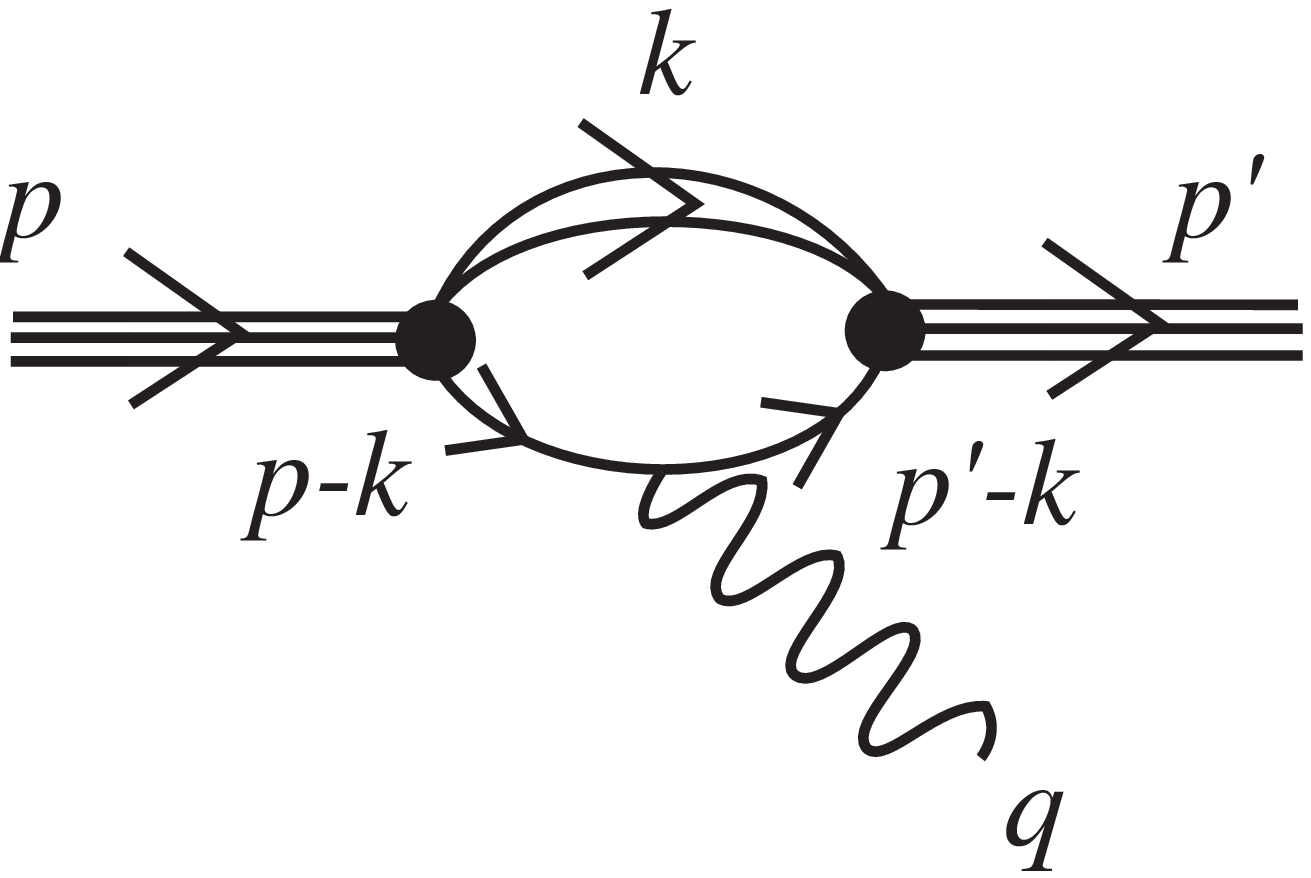}
\includegraphics[width=4cm]{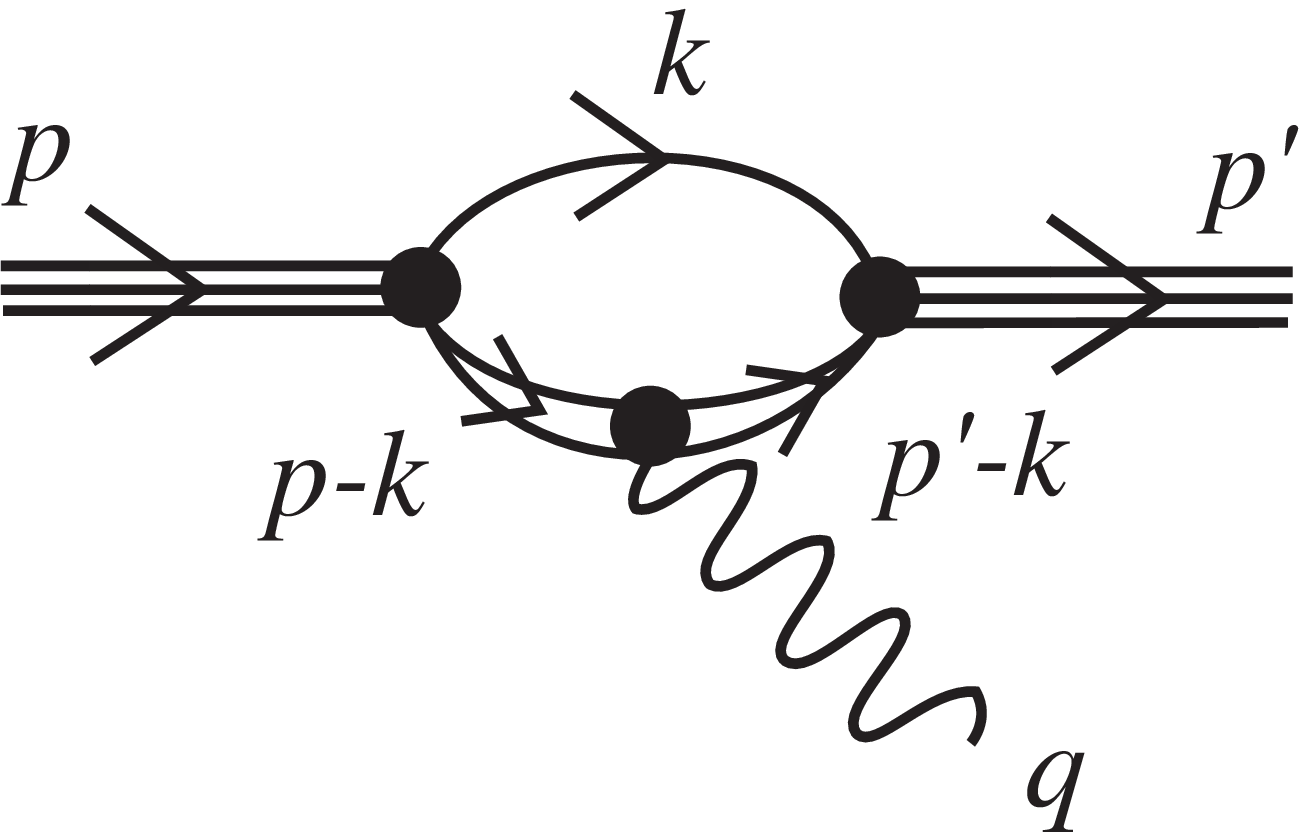}
}
\begin{minipage}{11cm}
   \caption{\small The electromagnetic
couplings of the baryons. The left and
right panels show the contribution from the quark and diquarks.}
   \label{fig:bbgamma}
\end{minipage}
\end{figure}

We proceed to the electromagnetic terms obtained from the first 
term of Eq.~(\ref{eq:trlgex}) and second one of Eq.~(\ref{eq:qex}),  
\begin{eqnarray}
{\cal L}_{BB\gamma}&=&-\bar{B}^\prime(p^\prime)\left(\begin{array}{cc}
\Lambda_{S}^\mu(q) & 0\\ 0 &
\Lambda_{A}^\mu(q) \end{array}\right)B^\prime(p) A_\mu(q),
\label{eq1}
\end{eqnarray}
where $\Lambda_S(q)$ and $\Lambda_A(q)$ are the electromagnetic form factors 
of $B_1^\prime$ and $B_2^\prime$, respectively. Each of them 
consists of the contributions from quark-photon and diquark photon 
couplings, as shown in Fig.~\ref{fig:bbgamma},
\begin{eqnarray}
\Lambda_S^\mu(q)&=&\Lambda_{Sq}^\mu(q)+\Lambda_{SD}^\mu(q),\\
\Lambda_A^\mu(q)&=&\Lambda_{Aq}^\mu(q)+\Lambda_{AD}^\mu(q).
\label{eq:decom}
\end{eqnarray}
Their Feynman integrals are given by,
\begin{eqnarray}
\Lambda_{Sq}^\mu (q) &=& iQ_q N_c Z_S
\int\frac{d^4k}{(2\pi)^4}\Delta_0(k)S_0(p^\prime-k)\gamma^\mu
S_0(p-k),\\
\Lambda_{SD}^\mu (q)&=& -i Q_S N_c Z_S
\int\frac{d^4k}{(2\pi)^4}\Delta_0(p^\prime-k)\Gamma^\mu_S(p,\;p^\prime)
\Delta_0(p-k)S_0(k),
\label{eq:emsclr}\\
\Lambda_{Aq}^\mu(q) &=& \frac32 i N_c Z_A(\tau_0-\tau_3)\int\frac{d^4k}{(2\pi)^4}
\Delta_0^{\alpha\beta}(k)\gamma_\alpha\gamma_5 S_0(p^\prime-k)\gamma^\mu
S_0(p-k)\gamma_\beta\gamma_5,\nonumber\\
 \\
\Lambda_{AD}^\mu(q) &=& 3 i  N_c Z_A (\tau_0+2\tau_3)\nn
\\ &\times& \int
\frac{d^4k}{(2\pi)^4}\Delta_{0\alpha\beta}(p^\prime-k)(-\Gamma_A^{\beta\gamma\mu}(p,\;p^\prime))\Delta_{0\gamma\delta}(p-k)\gamma^\alpha\gamma_5
S_0(k) \gamma^\delta \gamma^5.\nonumber \\
\label{eq:emaxial}%
\end{eqnarray}%
where $p$ and $p^\prime$ are the four-momentum of $B^\prime_{1,2}$ in 
the initial and final states, and $q=p^\prime-p$ is the four-momentum transfer carried
by the photon.
Note that each  $B^\prime_{1,2}$ is an isodoublet having two components: 
the charge $+1$ and $0$.
Then $\tau_0$ and $\tau_3$ describe the iso-scalar and iso-vector form factors.

For the computation of the form factors at finite momentum transfer 
$q\neq 0$, we employ the Breit frame defined by
$p^\prime=(p_0,\;\vec{q}/2),\;p=(p_0,\;-\vec{q}/2),\;q=(0,\vec{q})$.
The form factors $\Lambda^\mu_\alpha,\; (\alpha=Sq,\; SD,\; Aq,\; AD)$ 
are decomposed into the electric form factors 
$\Lambda^0_\alpha(q)=\Lambda^E_\alpha({\vec{q}}^{\,2}) \gamma^0$, 
and the magnetic form factors
$\Lambda^i_\alpha(q) =\Lambda^M_\alpha({\vec{q}}^{\,2}) i\epsilon_{ijk}\sigma_j q_k\; (i=1,\;2,\;3)$.
Then Eq.~(\ref{eq1}) is reduced to
\begin{eqnarray}
{\cal L}_{BB\gamma}=&-&\bar{B}^\prime(p^\prime)\left(\begin{array}{cc}
\Lambda_S^E({\vec{q}}^{\,2}) & 0 \\ 0 & \Lambda_A^E({\vec{q}}^{\,2})\end{array}\right)\gamma^0
B^\prime(p) A_0\nonumber\\
&-&\bar{B}^\prime(p^\prime)\left(\begin{array}{cc}
\Lambda_S^M({\vec{q}}^{\,2}) & 0 \\ 0 &
\Lambda_A^M({\vec{q}}^{\,2})\end{array}\right)i\epsilon_{ijk}\sigma_j q_k
B^\prime(p) A_i.
\label{eq:emmtrx}
\end{eqnarray}
By the unitary transformation Eq.(\ref{eq:unitary}), we obtain 
the electromagnetic couplings of the nucleon and Roper resonance as
\begin{eqnarray}
{\cal L}_{NN\gamma}=-\bar{N}(p^\prime)F^E({\vec{q}}^{\,2})\gamma^0
N(p) A_0-\bar{N}(p^\prime)F^M({\vec{q}}^{\,2})i\epsilon_{ijk}\sigma_j q_k
N(p) A_i,
\end{eqnarray}
where $F^{E,M}$ are defined by the unitary transformation of  Eq.~(\ref{eq:emmtrx}):
\begin{eqnarray}
F^{E,M}({\vec{q}}^{\,2})&=&
U \left(\begin{array}{cc}
\Lambda_S^{E,M}(\vec{q}^{\,2}) & 0 \\ 0 &
\Lambda_A^{E,M}(\vec{q}^{\,2})\end{array}\right)U^\dagger _,\nonumber\\
&=&\left(\begin{array}{cc}
\Lambda_S^{E,M}\cos^2\phi+\Lambda_A^{E,M}\sin^2\phi &
(\Lambda_A^{E,M}-\Lambda_S^{E,M})\sin\cos\phi \\
(\Lambda_A^{E,M}-\Lambda_S^{E,M})\sin\cos\phi & 
\Lambda_S^{E,M}\sin^2\phi+\Lambda_A^{E,M}\cos^2\phi
\end{array}\right)_{.}
\label{eq2}
\end{eqnarray}
Eq.~(\ref{eq2}) is the $2\times 2$ matrix in the nucleon-Roper space, 
and each element is also a $2\times 2$ matrix in  
iso-doublet space.
Therefore, Eq.~(\ref{eq2}) describes the electromagnetic properties of the four 
particles: the nucleon ($p$ and $n$), and the Roper resonance ($p^*$ and $n^*$),  
where $p$ and $p^*$ have the charges $+1$ and $n$ and $n^*$ have $0$.
Thus the diagonal elements describe the form factors of 
them, while the off-diagonal ones their transitions. 
In the present paper, we shall study the diagonal components. 
The off-diagonal terms should be calculated in other frame, for 
instance the Roper-rest-frame.

It is convenient to redefine the form factors  as
\begin{eqnarray}
\Lambda_{Sq}^\mu(q) &\to& Q_q \Lambda_{Sq}^\mu,\\
\Lambda_{SD}^\mu(q) &\to& Q_S \Lambda_{SD}^\mu,\\
\Lambda_{Aq}^\mu(q) &\to& \frac32 (\tau_0-\tau_3) \Lambda_{Aq}^\mu,\\
\Lambda_{AD}^\mu(q) &\to& 3 (\tau_0+2\tau_3)\Lambda_{AD}^\mu.
\end{eqnarray}%
Then, the electric form factors of the proton and neutron are given as,
\begin{eqnarray}
G_p^E({\vec{q}}^{\,2})&=&\left[\frac23 \Lambda^E_{Sq}({\vec{q}}^{\,2})+\frac13 \Lambda^E_{SD}({\vec{q}}^{\,2})\right]\cos^2\phi
+\left[0\cdot \Lambda^E_{Aq}({\vec{q}}^{\,2})+ \Lambda^E_{AD}({\vec{q}}^{\,2})\right]\sin^2\phi,\nonumber\\
 \\
G_n^E({\vec{q}}^{\,2})&=&\left[-\frac13 \Lambda^E_{Sq}({\vec{q}}^{\,2})+\frac13 \Lambda^E_{SD}({\vec{q}}^{\,2})\right]\cos^2\phi
+\left[\frac13 \Lambda^E_{Aq}({\vec{q}}^{\,2})-\frac13 \Lambda^E_{AD}({\vec{q}}^{\,2})\right]\sin^2\phi.\nonumber \\
\label{eq:gep}
\end{eqnarray}
Those of the Roper resonance are obtained by 
the substitution of $\sin\phi \leftrightarrow \cos\phi$.
The magnetic form factors of the proton and neutron are given by,
\begin{eqnarray}
\frac{G_p^M({\vec{q}}^{\,2})}{2 M_N} &=& \left[\frac23 \Lambda^M_{Sq}({\vec{q}}^{\,2})+\frac13 \Lambda^M_{SD}({\vec{q}}^{\,2})\right]\cos^2\phi
+\left[0\cdot \Lambda^M_{Aq}({\vec{q}}^{\,2})+ \Lambda^M_{AD}({\vec{q}}^{\,2})\right]\sin^2\phi,\nonumber \\
 \\
\frac{G_n^M({\vec{q}}^{\,2})}{2M_N}&=&\left[-\frac13 \Lambda^M_{Sq}({\vec{q}}^{\,2})+\frac13 \Lambda^M_{SD}({\vec{q}}^{\,2})\right]\cos^2\phi
+\left[\frac13 \Lambda^M_{Aq}({\vec{q}}^{\,2})-\frac13 \Lambda^M_{AD}({\vec{q}}^{\,2})\right]\sin^2\phi. \nonumber \\
\label{eq:gmn}%
\end{eqnarray}
Again, those of the Roper resonance are obtained by the 
substitution of $\sin\phi\leftrightarrow\cos\phi$ and $M_N \to M_R$.

\subsection{Intrinsic Properties of Diquarks}
The form factors are defined with the point-like diquarks. 
It is natural that the diquarks have intrinsic structures.
First, the axial-vector diquark may have an anomalous magnetic moment, which
was already introduced in Eq.~(\ref{eq:anmls}). 
In addition both the diquarks have intrinsic form factors.
Expressing those intrinsic form factors by $\Lambda_{I,\alpha}({\vec{q}}^{\,2}),\; 
(\alpha=SD,\; AD)$, they can be included by multiplying to the original ones
\begin{eqnarray}
\Lambda_{\alpha}({\vec{q}}^{\,2}) \to \Lambda_{\alpha}({\vec{q}}^{\,2}) \Lambda_{I,\alpha}({\vec{q}}^{\,2}).
\end{eqnarray}
Here we introduce a common form factor for both the electric and magnetic
terms. Then we parameterize them as,
\begin{eqnarray}
\Lambda_{I, SD}({\vec{q}}^{\,2})=\frac{1}{(1+{\vec{q}}^{\,2}/b_S)},\\
\Lambda_{I, AD}({\vec{q}}^{\,2})=\frac{1}{(1+{\vec{q}}^{\,2}/b_A)^2},
\label{eq:idff}%
\end{eqnarray}%
We have employed the monopole and dipole shapes for 
the scalar and axial-vector diquarks, followed by Kroll et 
al.~\cite{Kroll:1993zx}. Those shapes are obtained 
 by considering the magnetic form factors of the nucleon
 in high energy region. 
Although it is not obvious that  the parameterization can be applied to 
 the low-energy region ($\leq 1$ GeV)  we are interested in,
 we shall employ  Eqs.~(\ref{eq:idff}) in order to decrease the number of parameters.
The values of $b_S$ and $b_A$ control the sizes of the  diquarks. Explicitly the radii of 
the diquarks are given by 
\begin{eqnarray}
\bra r^2\ket_S=\frac{6}{b_S}_,\\
\bra r^2\ket_A=\frac{12}{b_A}_.
\end{eqnarray}
\section{Numerical Results}
\label{sec:em}

\subsection{Masses and Parameters}
\label{sec:numerics}
For numerical calculations, we first explain the model parameters. 
The constituent mass of the $u,d$ quarks $m_q$ and the three momentum 
cutoff $\Lambda$ are determined in such a way that they 
reproduce meson properties in the NJL model~\cite{Vogl:1991qt,Hatsuda:1994pi}. 
The masses of the diquarks may be also calculated in the NJL
model~\cite{Vogl:1991qt,Cahill:1987qr}. 
We choose the mass of the scalar diquark $M_S$=650 MeV.
The mass difference $M_A - M_S$
is related to that of the nucleon and  $\Delta(1232)$.  
Hence, we assume the mass difference $M_A-M_S=400$ MeV with the mass
of the axial-vector diquark  $M_A=1050$ MeV~\footnote{
In the lattice QCD calculations, Orginos obtained  $M_A-M_S=360$ MeV~\cite{Orginos:2005vr}, while Hess et al. obtained  $M_A-M_S\sim 100$ MeV~\cite{Hess:1998sd}. Note that Hess
obtained  $M_\Delta-M_N\sim 180$ MeV.}.
The quark-diquark coupling constants $G_S,\;G_A$ and $v$ are determined  so as to 
reproduce the masses of the nucleon and Roper resonance~\cite{Nagata:2005qb}.
The parameters are summarized in Table~\ref{tab:parameter2}.
\begin{table}[tbh]
\begin{center}
\caption{The parameters used in this model.}
\begin{tabular}{ccccc}
\hline
$m_q$ $ [$GeV] & $M_S$ [GeV]& $M_A$ [GeV]& $m_N$ [GeV]& $m_R$ [GeV]\\
\hline
 0.39 & 0.65 & 1.05 & 0.94 & 1.44  \\
\hline
\end{tabular}
\begin{tabular}{ccccc}
\hline
 $\Lambda$ [GeV]&  $G_S$ [GeV$^{-1}$]& $G_A$ [GeV$^{-1}$]& $v$  [GeV$^{-1}$]& $\phi$ [degree]\\
\hline
 0.6 &  102 & 11.2 & 23.2 & 18.4 \\
\hline
\end{tabular}
\label{tab:parameter2}
\end{center}
\end{table}

In the previous section, we have introduced three parameters, $b_S$, $b_A$ and $\kappa$.
We employ $b_S=1.25$ [GeV$^2$] with the radius of the scalar diquark 
$\bra r^2\ket_S^{1/2}=0.50$ [fm] obtained in the NJL model~\cite{Weiss:1993kv}.
In Ref.~\cite{Weiss:1993kv}, the axial-vector diquark has the large size $1\sim 2$ [fm]. 
Rather we assume that the axial-vector diquark is  smaller than  the nucleon,   
and that it is larger than the scalar diquark, because there are no 
strong attraction between the quarks in the axial-vector diquark.
Hence the axial-vector diquark size is assumed to be within a range 
$\bra r^2\ket_A^{1/2}= 0.6\sim 0.9$ [fm].
Within this range, we determine  $\bra r^2\ket_A^{1/2}=0.82$ [fm] with $b_A=0.7$, which are appropriate to reproduce the charge radii of the proton and neutron~\footnote{Kroll et al. employed smaller diquark radii, or larger values of $b_S$ and $b_A$ rather than our's. Recent lattice calculation reported larger values of the scalar diquark size about $1$ [fm]~\cite{Alexandrou:2006cq}. It is not suitable to reproduce the charge radii of the proton 
and neutron in the present model setup. }. 

For the determination of $\kappa$,  it turns out that it is not easy to reproduce 
the magnetic moments of the nucleon with a reasonable value of $\kappa$, 
because of the scalar dominance with the small mixing angle $\phi\sim 18^\circ$. 
Then, we shall qualitatively investigate the effects of $\kappa$
to the magnetic moments with employing $\kappa=2$, which is larger than the values used in Kroll et al. 

\subsection{Electromagnetic Properties}

\begin{figure}[tbh]
\centering{
\includegraphics[width=5cm]{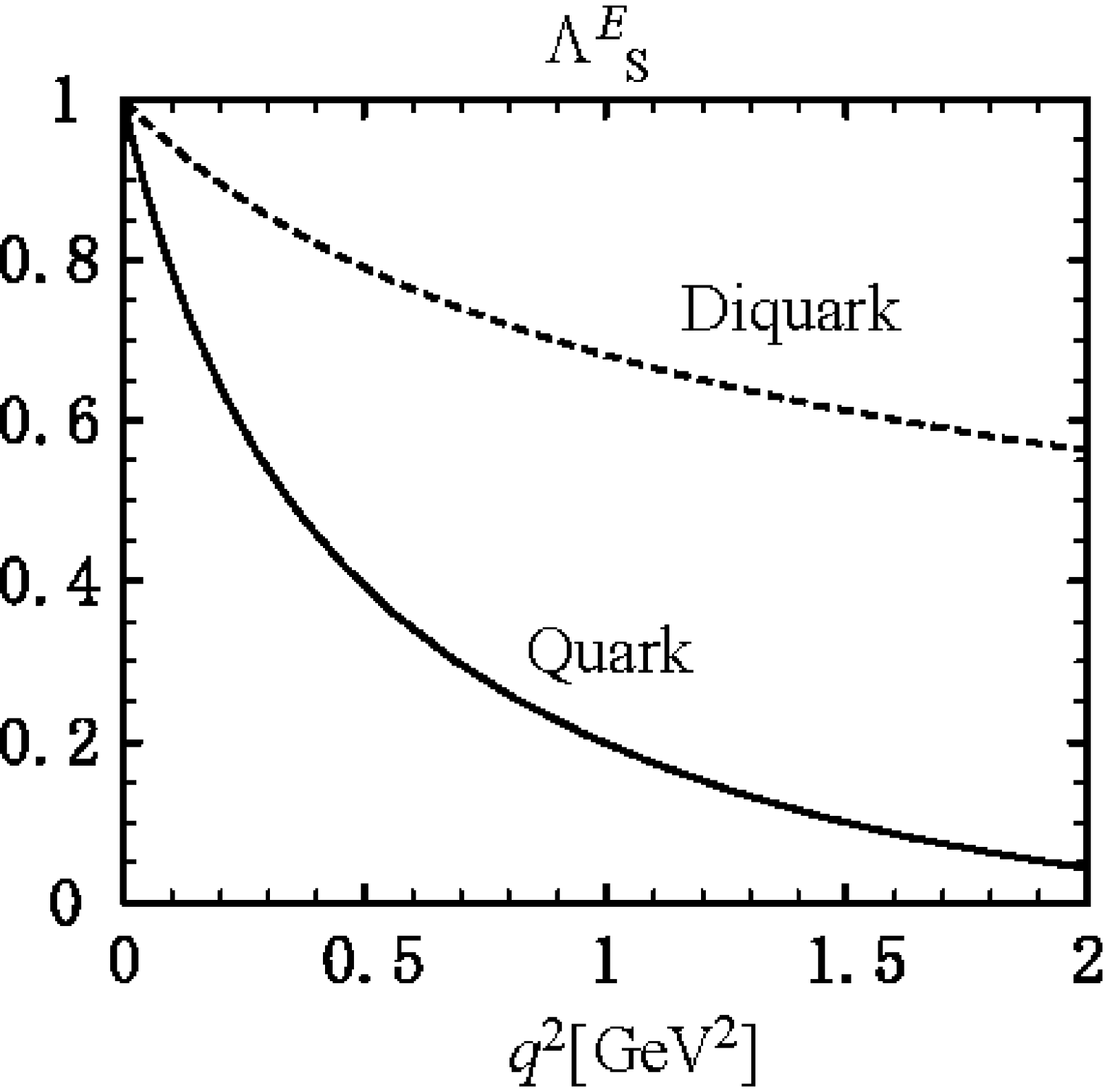}
\includegraphics[width=4.7cm]{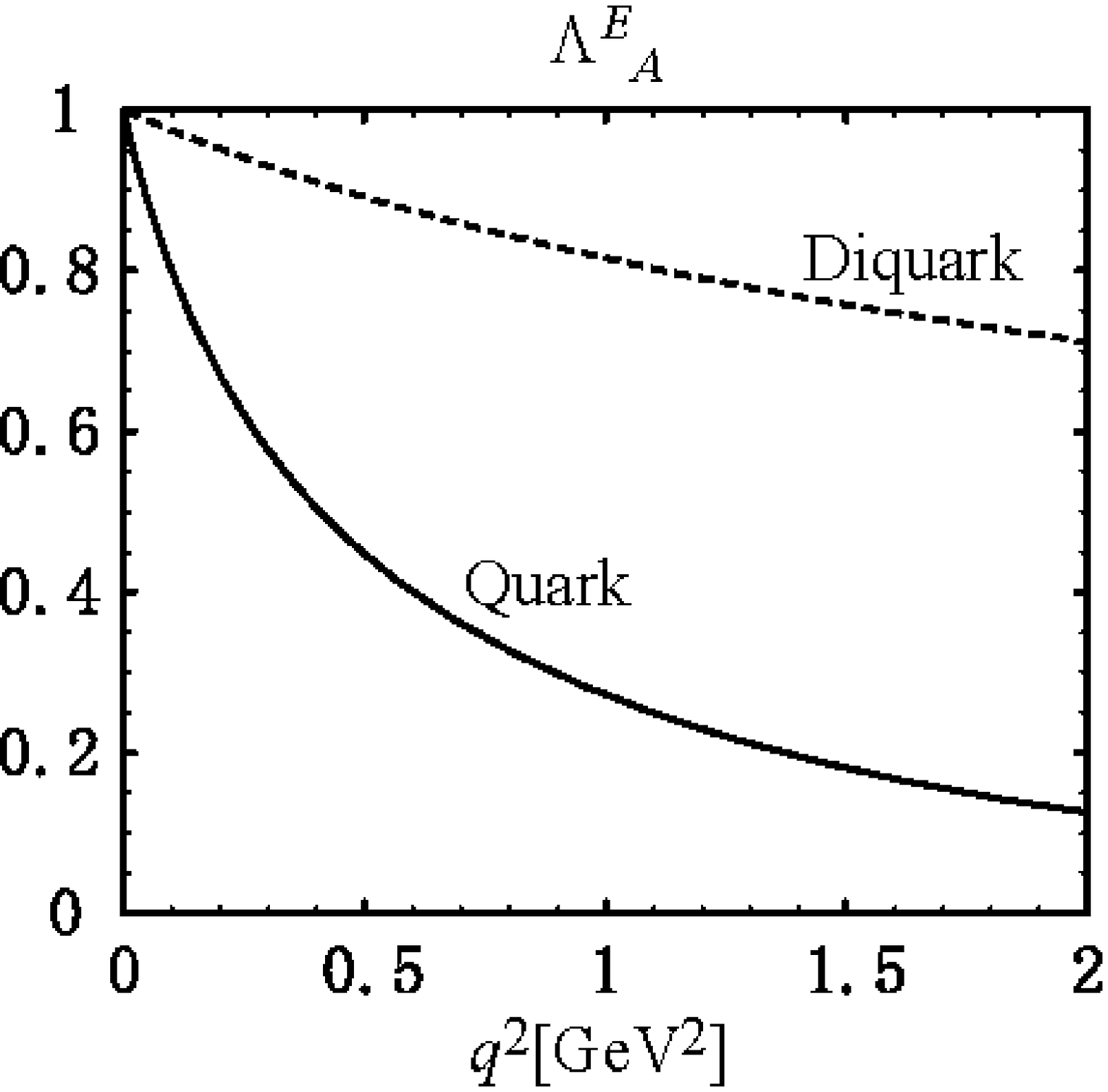}
\includegraphics[width=5cm]{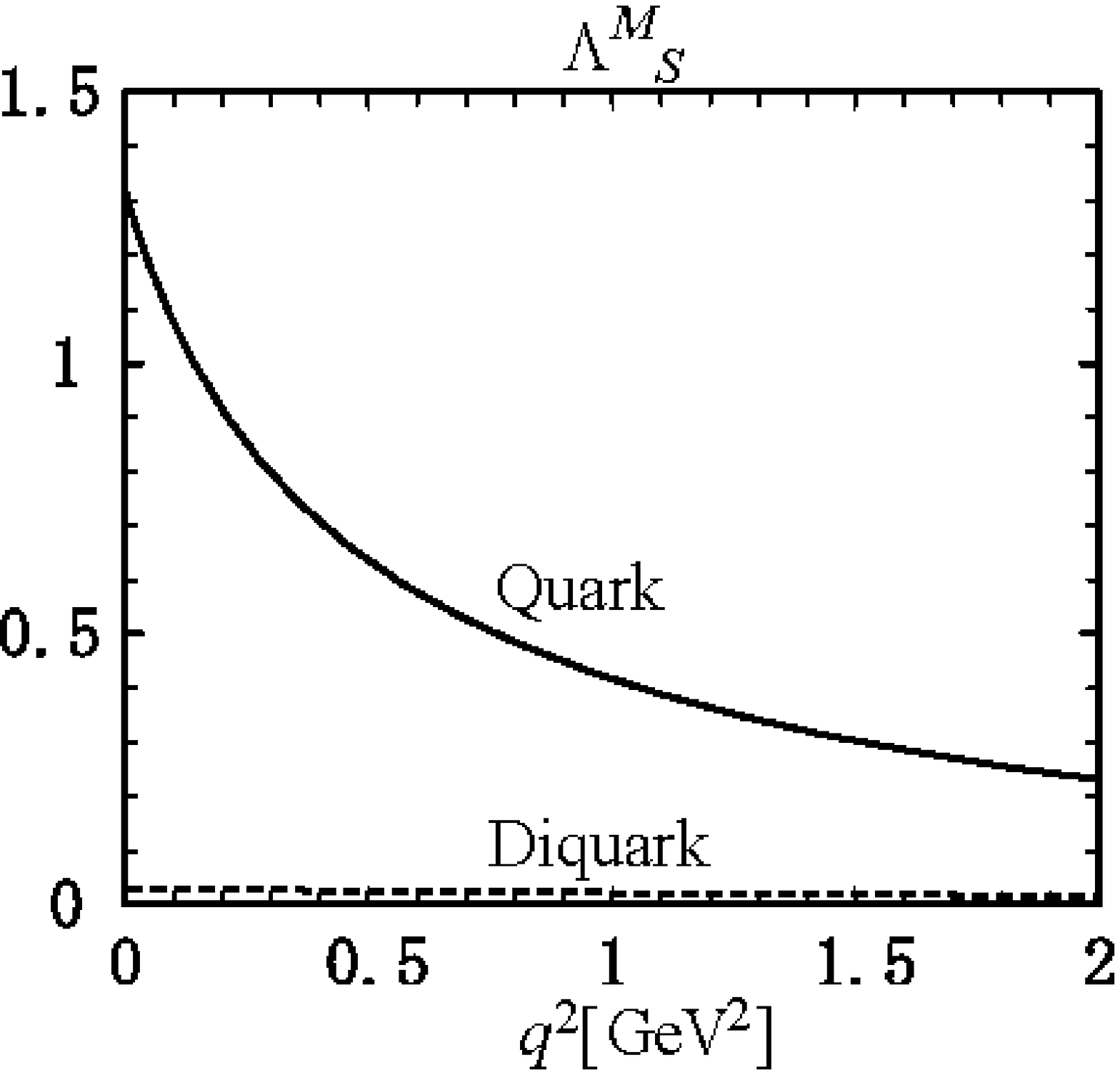}
\includegraphics[width=5cm]{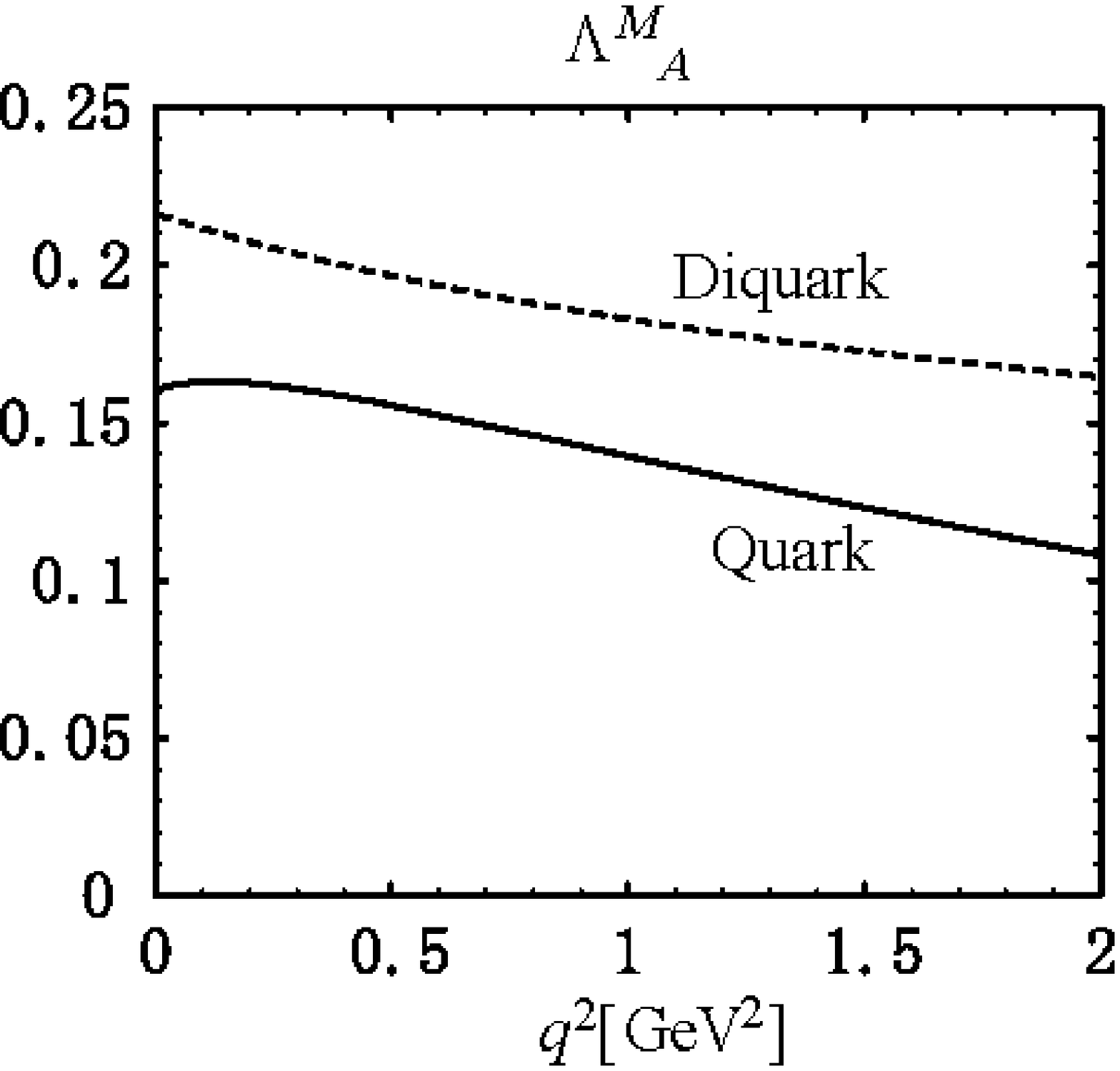}
}
\begin{minipage}{10cm}
\caption{\small The eight fundamental form factors. 
The topped and bottomed figures show electric and magnetic form factors, 
where the left and right panels show  the scalar and axial-vector channels.
 The solid and dashed lines are for the quark and diquark contributions.}
\label{fig:8ff}
\end{minipage}
\end{figure}

The top panels of Fig.~\ref{fig:8ff} show the electric form factors 
 $\Lambda^{E}_{Sq,Aq,SD,AD}$. Here we do not include the intrinsic
diquark form factors, hence the figures show the distributions 
of the orbital motion of the quark and diquarks.
In both the scalar (left panel) and the axial-vector (right) channels, the quark 
contributions $\Lambda^E_{Sq,Aq}$ have larger
$\vec{q}^{\;2}$-dependence than the diquarks $\Lambda^E_{SD,AD}$.  
In addition, the quarks have almost the same distributions in the scalar and 
axial-vector channels.
This is understood from the system of a lighter quark and a heavier diquark, 
where the quark moves further from the center of mass than the diquarks. 
Since the quark-scalar diquark and quark-axial-vector diquark bound states have 
the same binding energy 50 MeV, 
the quarks have the similar spatial distributions and  kinetic energies.
As for the diquark contributions, the scalar diquarks have 
larger $\vec{q}^{\;2}$-dependence than the axial-vector diquarks, however both 
of them are almost negligible.
The bottom panels of Fig.~\ref{fig:8ff} show the magnetic form factors of 
$\Lambda^{M}_{Sq,Aq,SD,AD}$.
The magnetic moments are dominated by $F^M_{Sq}$, while the 
other contributions are rather small, which is consistent with 
the Faddeev equation by Mineo et. al~\cite{Mineo:2002bg}.

\begin{figure}[hbt]
\centering{
\includegraphics[width=5cm]{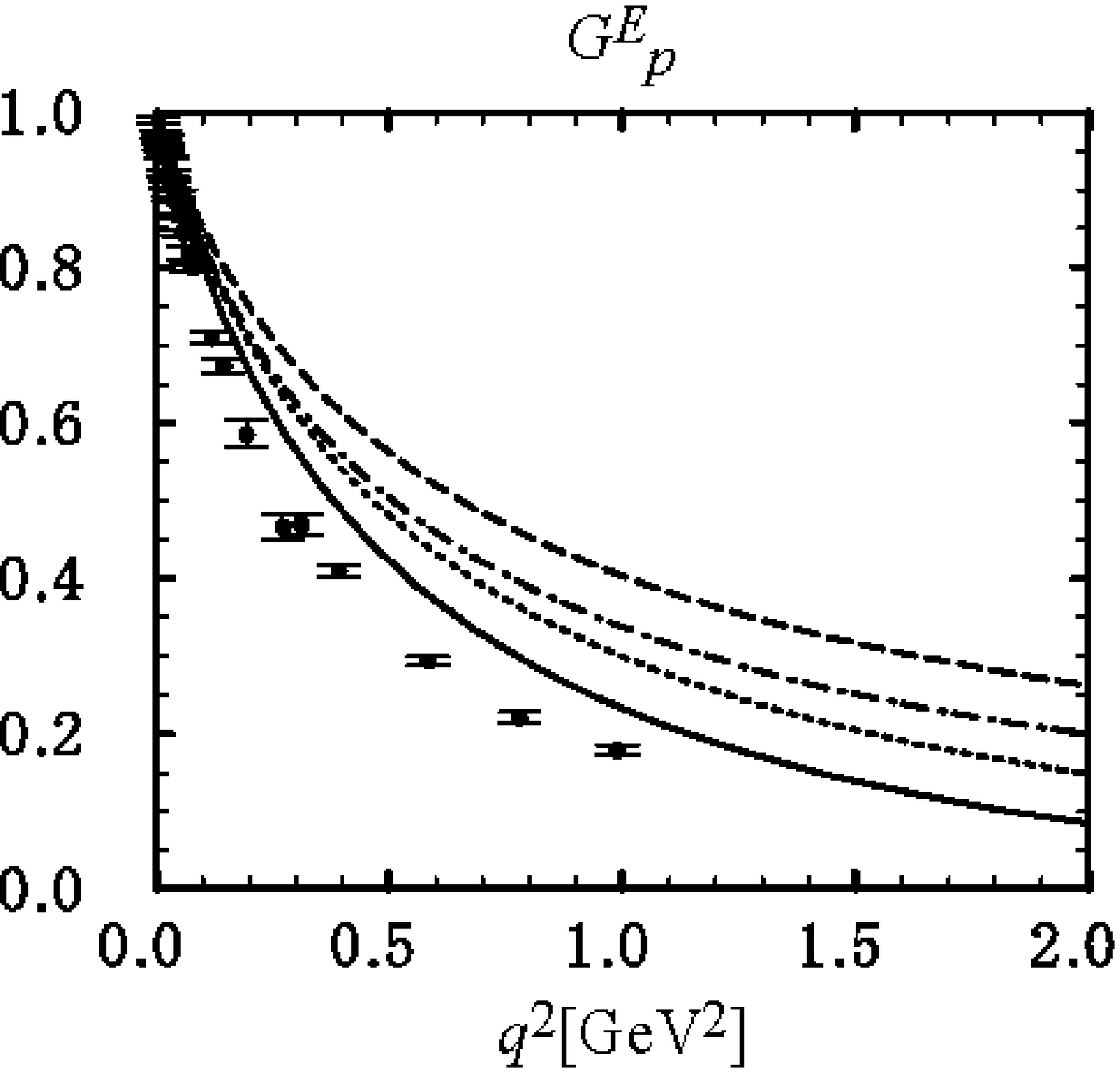}
\includegraphics[width=5.5cm]{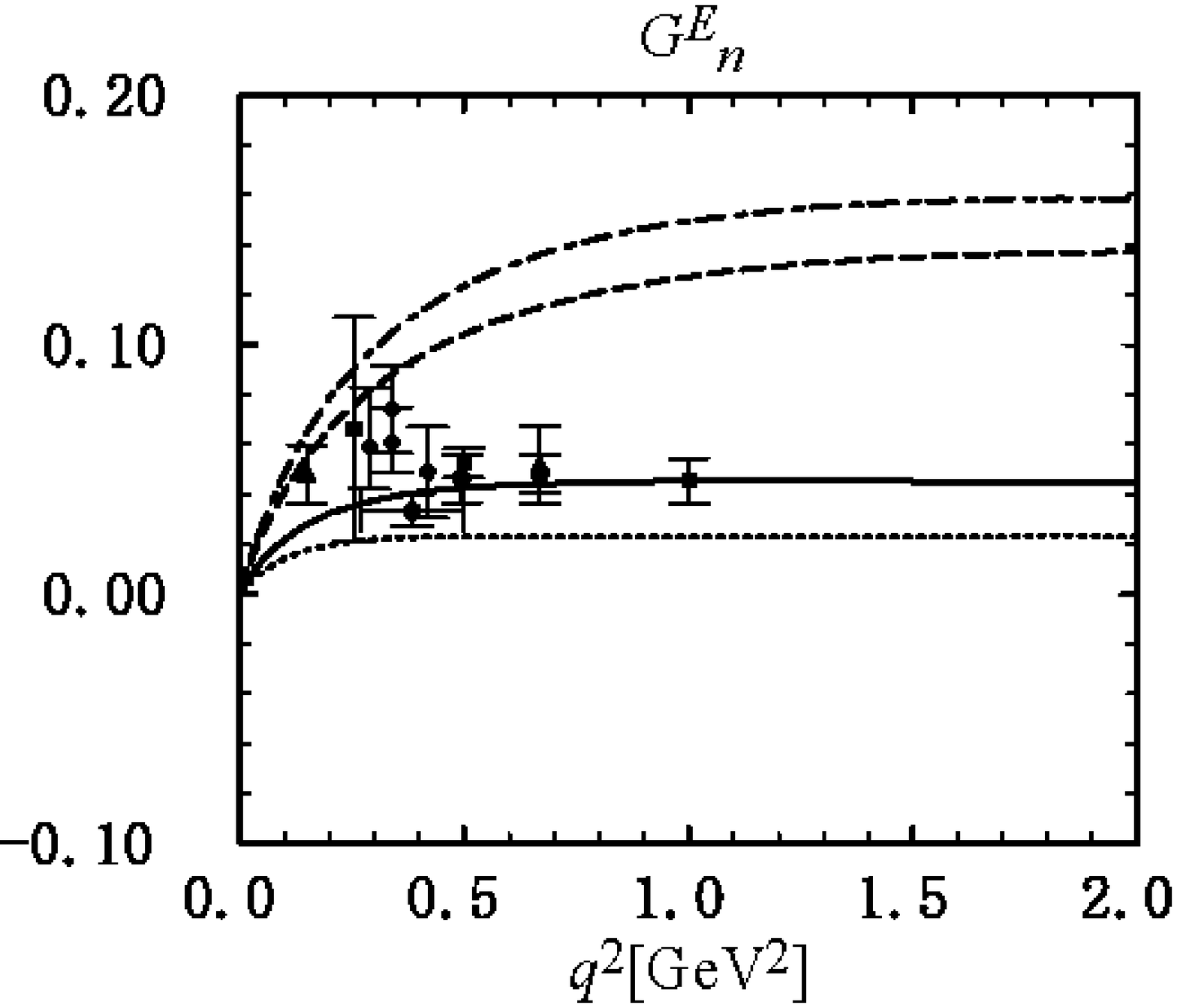}}
\begin{minipage}{11cm}
\caption{\small The electric form factors of the proton (left) and neutron
(right). The dotted, dotted-dashed, solid and dashed lines are for the cases (i), (ii), (iii) and (iv).  
The experimental data are taken 
from~\protect\cite{Hohler:1976ax,Eden:1994ji}.}\label{fig:idff}%
\end{minipage}%
\end{figure}%
Figures~\ref{fig:idff} show the electric form factors of the proton and neutron. 
To see the effects of the intrinsic diquark form factors, we consider 
the following four cases; case (i) include the intrinsic form factor only for the 
scalar diquark, case (ii) only for the axial-vector 
diquark, case (iii) for both the diquarks and case (iv) for
neither the scalar nor axial-vector diquarks. 
The intrinsic form factor of the scalar diquark 
improves both $G^E_p$ and $G^E_n$, while
that of the axial-vector diquark improves $G^E_p$ but  $G^E_n$. 
Including both the scalar and axial-vector diquark form factors, we obtain 
the reasonable results both for $G^E_p$ and $G^E_n$ in 
the region $0\leq \vec{q}^{\;2} \leq 1$ [GeV$^2$].

\begin{figure}[hbt]
\centering{
\includegraphics[width=5cm]{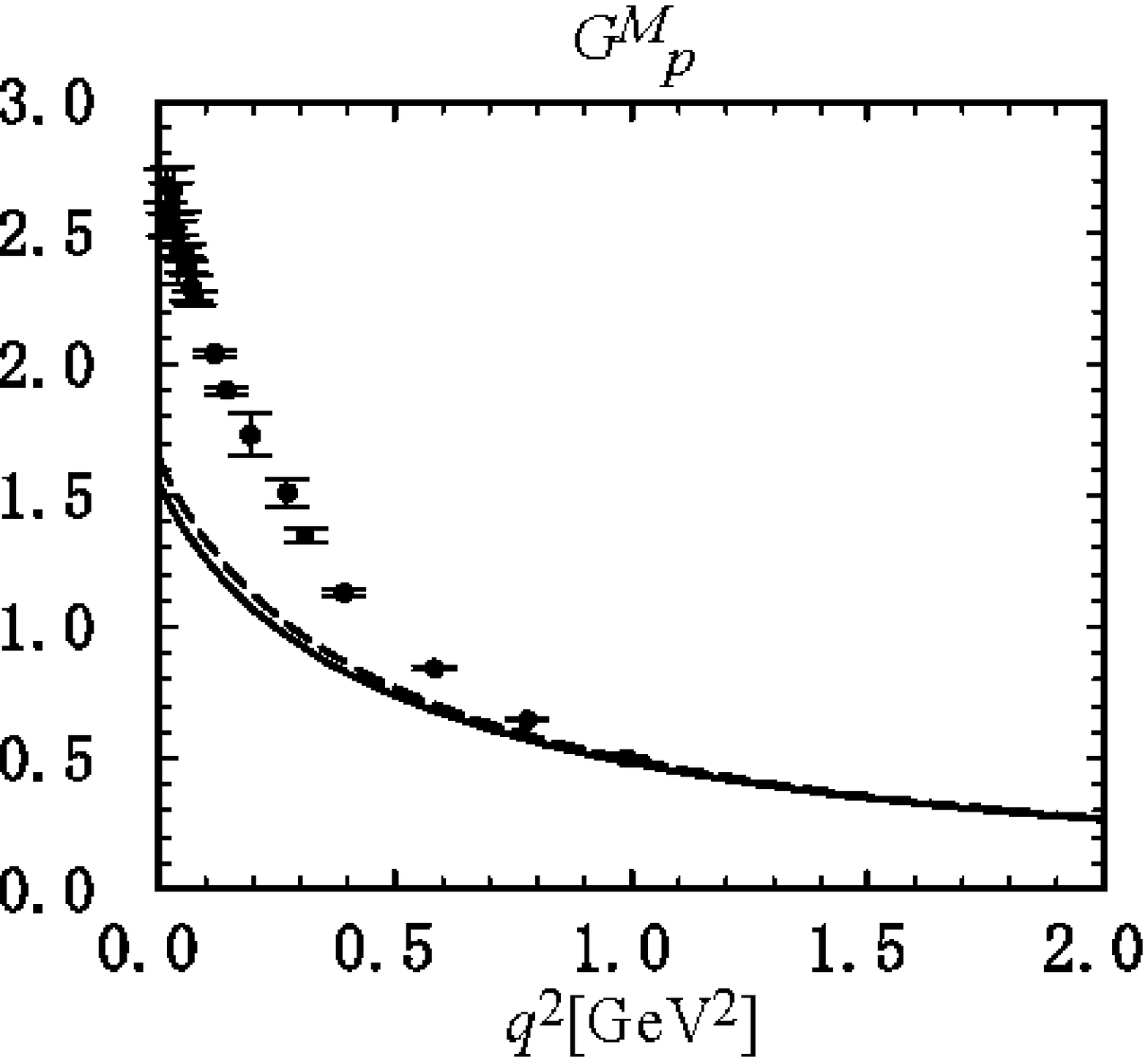}
\includegraphics[width=5.5cm]{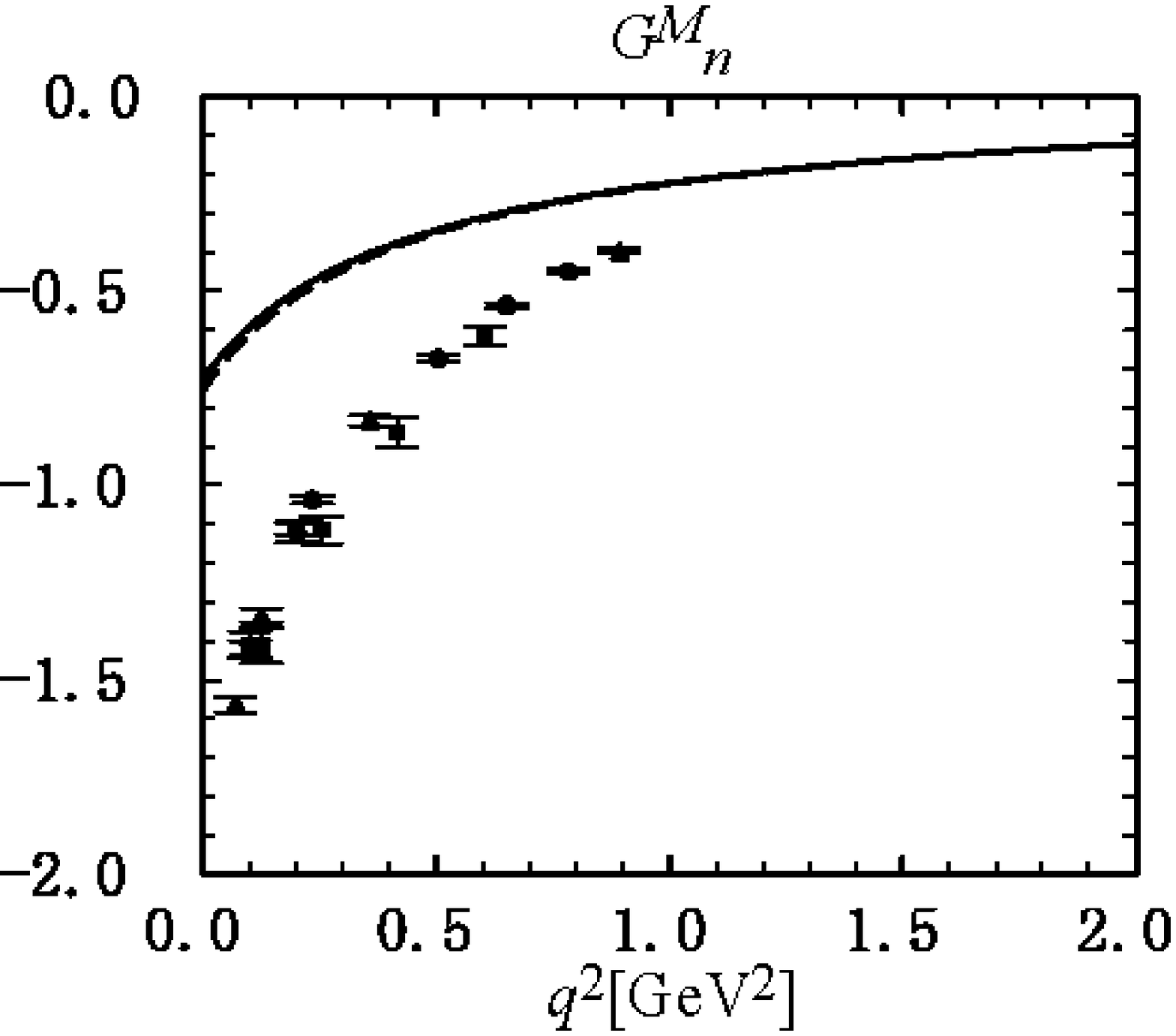}}
\begin{minipage}{11cm}
\caption{\small The magnetic form factors of the proton (left) and
neutron (right).   The solid and dashed lines are for $\kappa=0$ and $\kappa=2$. 
The experimental data are taken 
from~\protect\cite{Hohler:1976ax,Bruins:1995ns}.}\label{fig:kappa_dep_gm}
\end{minipage}
\end{figure}
Figures~\ref{fig:kappa_dep_gm}  plot the magnetic form factors of the proton and neutron.
Here we qualitatively estimate the effects of the anomalous magnetic moment
$\kappa$ of the axial-vector diquark by considering two cases; one is a result 
with $\kappa=0$ and the other is with $\kappa=2$.
For both cases, the intrinsic diquark form factors are included. 
We find that the magnetic moments of the proton and neutron are 
underestimated even if we include the anomalous magnetic moment:
$\mu_p=1.6$ (exp. $2.8$), $\mu_n=-0.76$ $(-1.9)$ for $\kappa=2$. 
The deficiency is mostly found in the iso-vector magnetic moments: 
$\mu_S=0.42$ (exp. $0.44)$ and $\mu_V=1.2$ $(2.4)$. 
It is caused by  the ratio of the weight factors 
$\cos^2\phi:\sin^2\phi=9:1$ due to the small mixing angle $\phi=18^\circ$.
Since the magnetic moments of the nucleon are 
dominated by the scalar channels, the anomalous magnetic 
moment of the axial-vector diquark does not contribute $\mu_{p,n}$ largely~\footnote{
Szweda et. al.~\cite{Szweda:1996wb} and 
Hellstern et al.~\cite{Hellstern:1997pg} showed that the magnetic moments 
are reproduced with the smaller value of the axial-vector diquark mass.
In our model, we can not employ a small mass for the diquark because
of the lack of confinement.}.

Mineo et al. showed that the electromagnetic
transition vertex between the scalar and axial-vector diquarks enhanced
the magnetic moments around $2.3$ for proton and $-1.5$ for neutron. 
They also showed that the remaining deficiencies were supplied by 
the effect of the pion~\footnote{The importance of the pion has been extensively studied, 
for instance see Refs.~\cite{Hosaka:1996ee,Thomas:1982kv}.}.
Similar results were also obtained in the covariant 
quark-diquark model by Oettel et al.~\cite{Oettel:2000jj}.
We can expect that  the transition vertex contributes the magnetic moments largely, 
because the contribution of the transition vertex appear  with a factor 
$\sin\phi\cos\phi$. 
In addition, the transition vertex contributes only to the iso-vector magnetic moment,
where the deficiency is observed in our results.

\begin{figure}[hbt]
\centering{
\includegraphics[width=5cm]{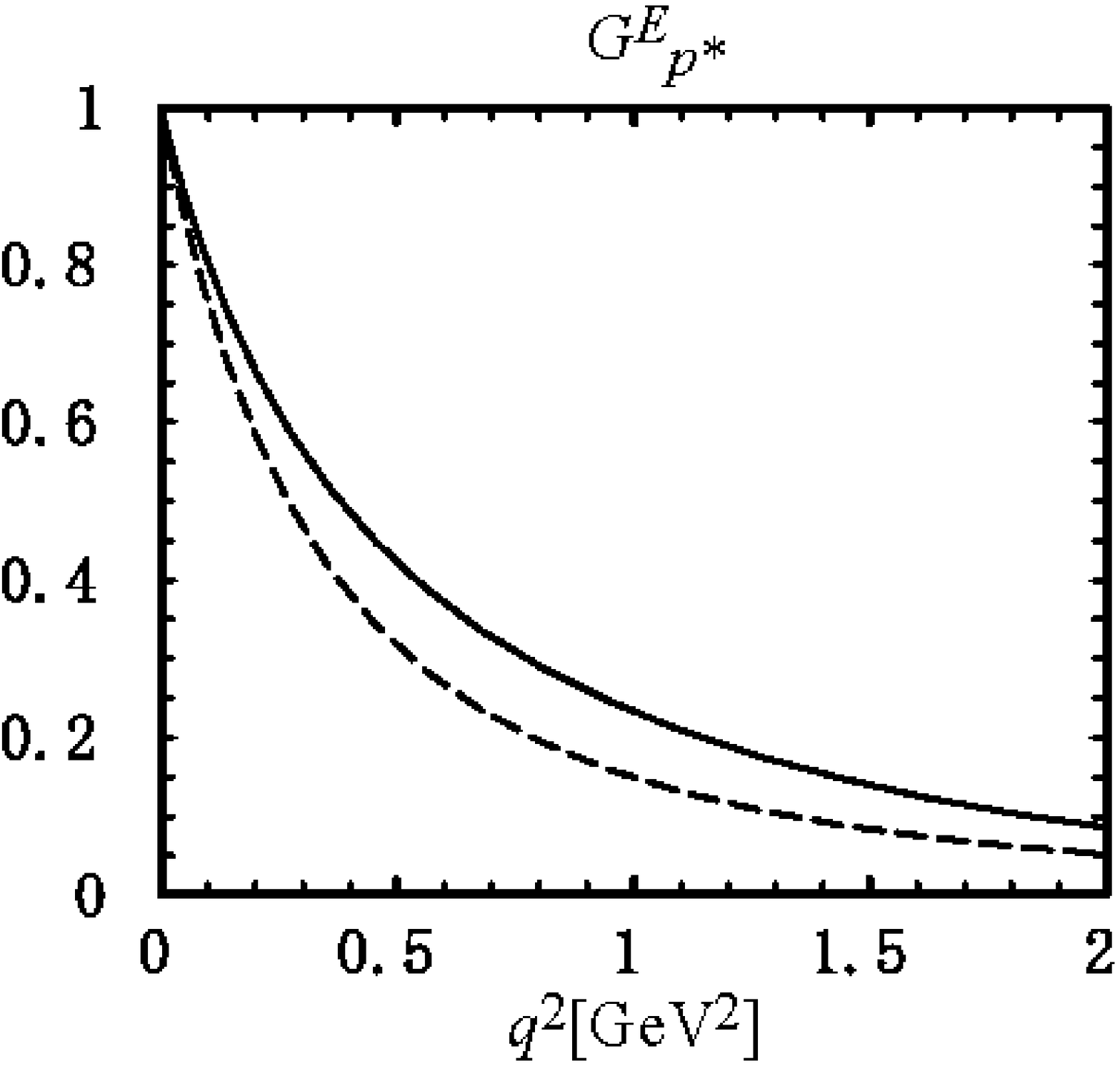}
\includegraphics[width=5.15cm]{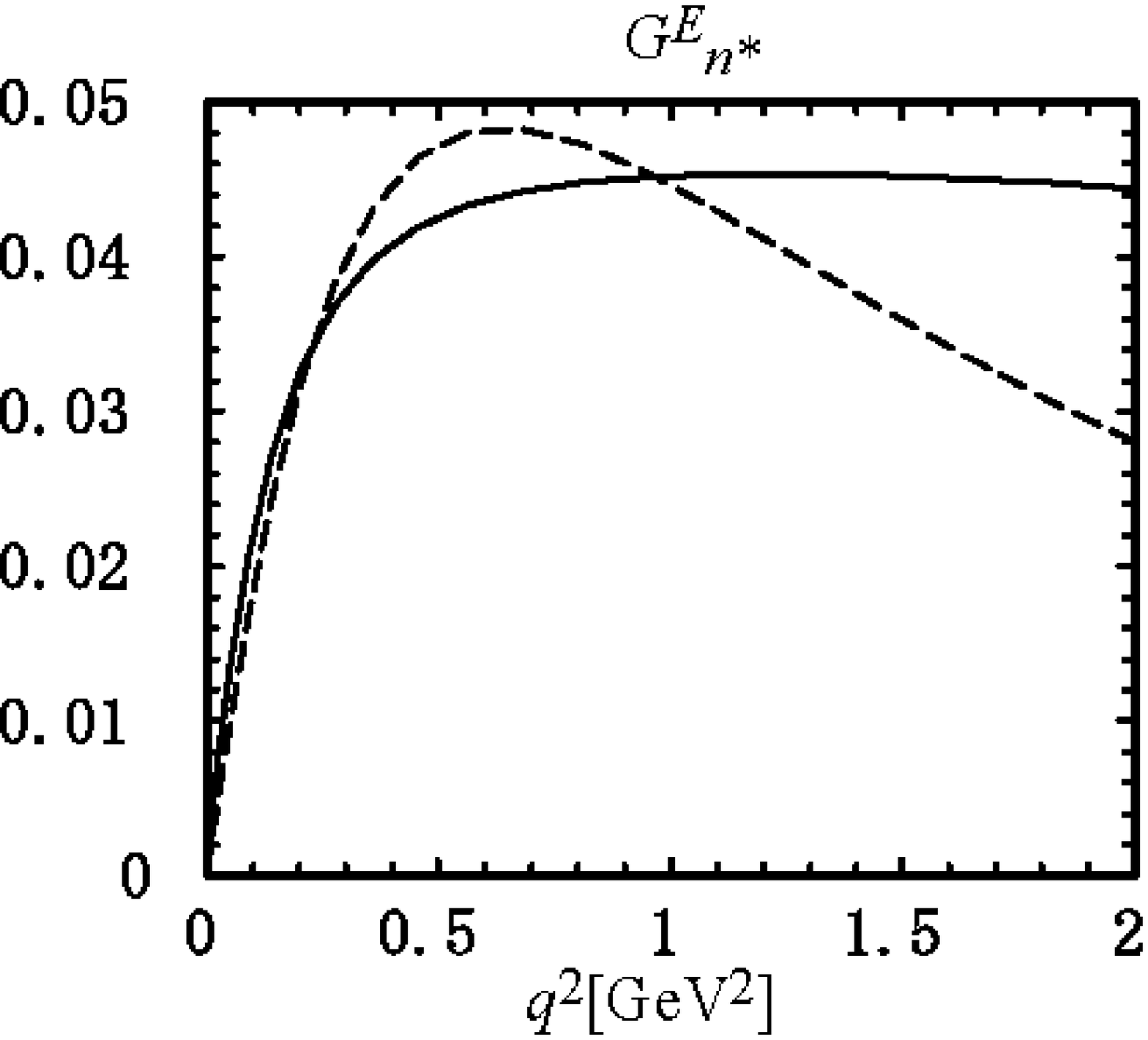}}
\begin{minipage}{11cm}
\caption{\small The electric form factors of the Roper resonance,  $p^*$ (left) and  $n^*$ (right). 
For a reference, the dashed lines are for the 
proton and neutron. We employ $b_S=0.96,\; b_A=0.7$ [GeV$^2$].}
\label{fig:ge_NR}
\end{minipage}
\end{figure}
Now we proceed to the Roper resonance.
Figures~\ref{fig:ge_NR} show the electric form factors of the Roper resonance. 
The slope of $p^*$ at $\vec{q}^{\;2}=0$ is comparable to that of the proton, 
precisely the slope of $p^*$ is slightly larger than that of $p$.
Therefore the charge radius of $p^*$ is larger than that
of $p$. This is understood from that the charge radius of $p$
is dominated by the orbital motion of the quark in the scalar 
channel, while that of $p^*$ is dominated by the intrinsic size of 
the axial-vector diquark owing to the small mixing angle and
negligible orbital motion of the diquarks. 
For the neutron component (right panel), 
the charge radius of  $n^*$ are almost same value as that of $n$. 
For the charge radius of $n$, the orbital motion of quark with the charge 
$+1/3$ and the intrinsic size of the scalar diquark with $-1/3$ almost cancel, 
however the former is slightly larger than the latter.
Owing to this cancellation, the charge radius of $n$ become negative.
There is also similar cancellation for the charge radius of $n^*$.
In this case, however, the intrinsic size of the axial-vector diquark is slightly larger 
than the orbital motion of the quarks.  Hence the charge radius of $p^*$ becomes negative.
We note that the charge radius of $n^*$ is sensitive to the intrinsic 
size of the axial-vector diquark, as shown in Table~\ref{tab:idffAdep}, which is also 
understood from the cancellation mechanism explained above.
\begin{table}[hbt]
\begin{center}
\caption{The charge radii
for various values of $b_A$, where  $b_S$=0.96 [GeV$^2$].}\label{tab:idffAdep}
\begin{tabular}{cccccc}
\vspace*{-9mm}\\
\noalign{\hrule height 0.8pt}
$b_A$ [GeV$^2$]  & $\bra r^2\ket^{1/2}_p$ [fm] &  $\bra r^2\ket^{1/2}_{p^*}$ [fm] 
&  $\bra r^2\ket_n$ [fm$^2$] & $\bra r^2\ket_{n^*}$ [fm$^2$] & $\bra r^2 \ket^{1/2}_A$ [fm]\\
\hline
$0.7$ & 0.75  & 0.84 & -0.074 & -0.046 & 0.82 \\
$1$   & 0.74  & 0.73 & -0.067 & 0.013  & 0.68\\
$1.3$ & 0.73  & 0.68 & -0.065 & 0.036  & 0.60\\
\hline
\end{tabular}
\end{center}
\end{table}

\begin{figure}
\centering{
\includegraphics[width=4.7cm]{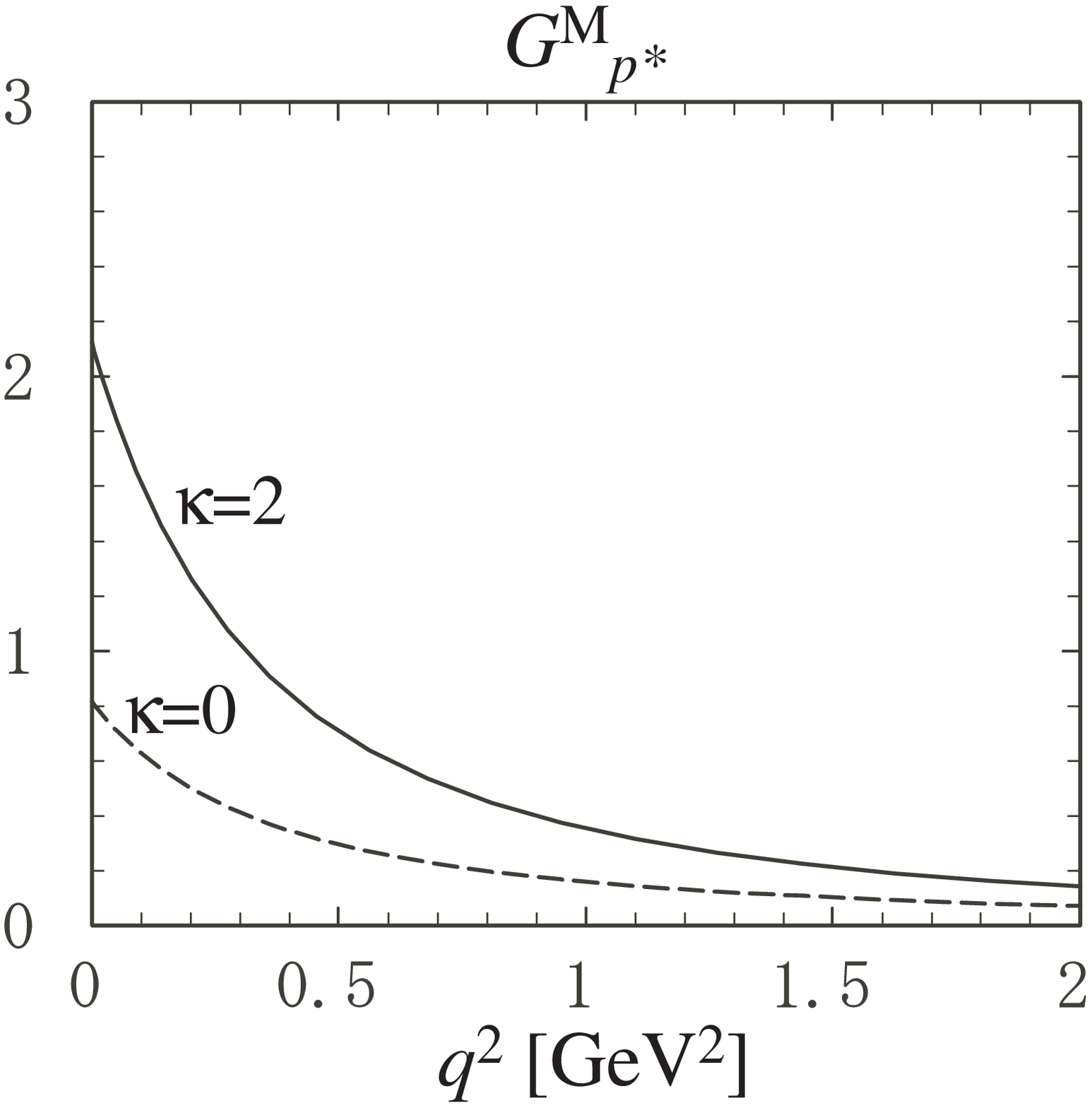}
\includegraphics[width=5cm]{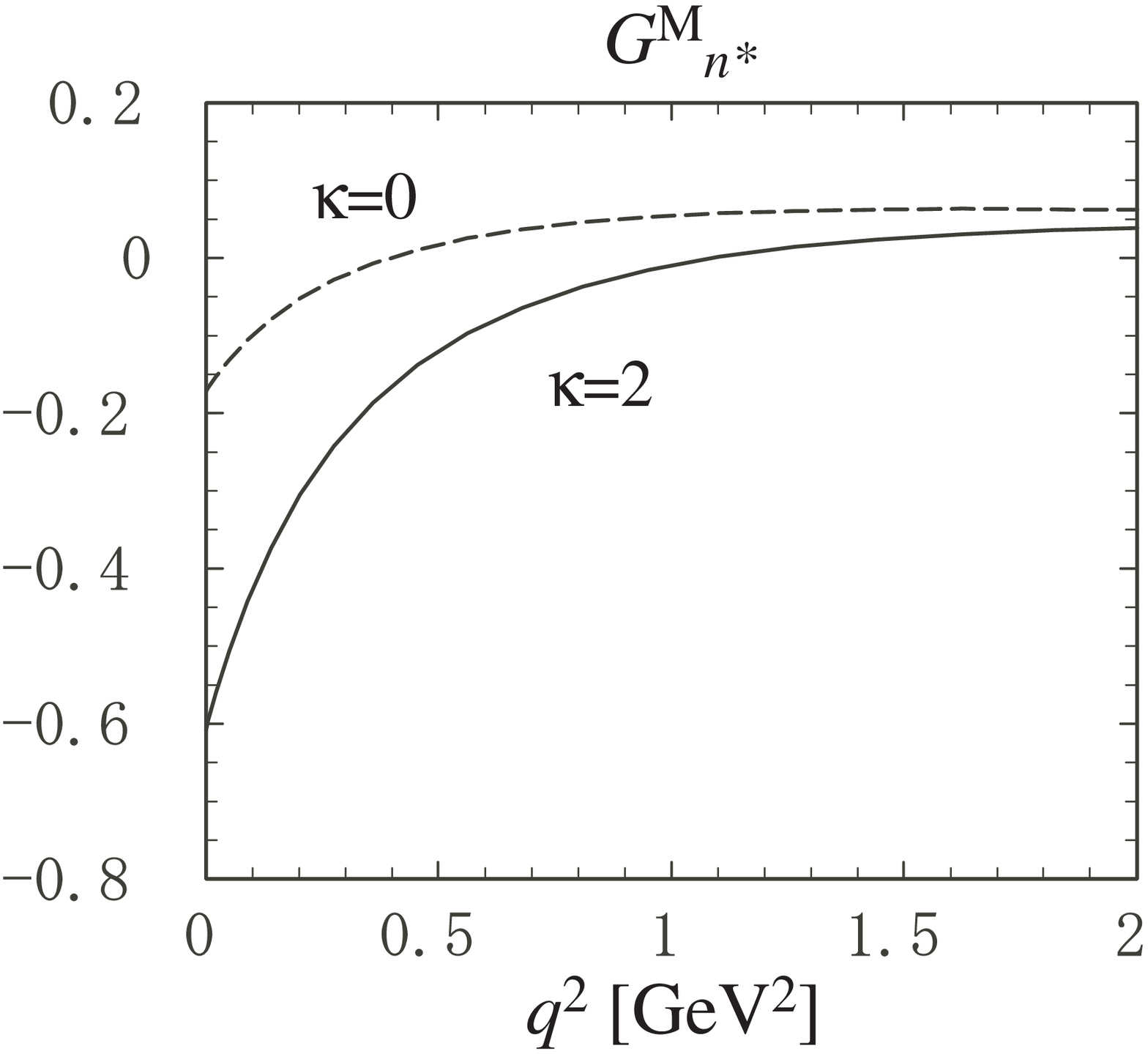}}
\begin{minipage}{11cm}
\caption{\small The magnetic form factors of the Roper resonance,  
$p^*$ (left) and  $n^*$ (right).
The solid and dashed lines are for $\kappa=2$ and $\kappa=0$.
 We employ $b_S=0.96,\; b_A=0.7$ [GeV$^2$].}
\label{fig:gm_NR}
\end{minipage}
\end{figure}
Figures~\ref{fig:gm_NR} show the magnetic form factors of the Roper 
resonance. We also consider the following two cases: $\kappa=0$ and $\kappa=2$.
The magnetic moments of the Roper are affected largely by 
the value of $\kappa$, in contrast to the case of the nucleon,
as is easily understood from the axial-vector dominance 
of the Roper resonance.
In the case $\kappa=0$, the magnetic moments of the Roper resonance is 
smaller than those of the nucleon, while in the case $\kappa=2$ 
the strengths of them are reversed. 
Now, we discuss the effects of $\kappa$. As we have explained 
we can obtain larger values of $\mu_{p,n}$ by employing the 
larger values of $\kappa$. In this case the magnetic moments
of the Roper become extremely large. Hence, we can not
determine the value of $\kappa$ at this moment, because there are no 
experimental data for the magnetic moments of the Roper resonance.
Again, we expect several effects for the magnetic moments: the scalar-axial-vector
diquark transition vertex, the pionic effect and so on.
The determination of $\kappa$ should be done by including such effects 
and by reproducing the proton and neutron magnetic moments.
Qualitatively, note that the magnetic moments of the nucleon and Roper 
can have different values. 
In the radial or collective excitation picture, they have the same amount of the 
magnetic moments, since the nucleon
and Roper resonance have the same flavor wave-function. 
In the quark-diquark picture, the Roper has different diquark components from
the nucleon. Hence the magnetic moments of the Roper differ from those of
the nucleon by the effects of the (difference of) diquark correlations.
Finally, Table~\ref{tab:mu_NR} summarize the properties of the nucleon and Roper resonance.
\begin{table}[hbt]
\begin{center}
\caption{The charge $Q$, magnetic moment $\mu$ and 
charge radius $\bra r^2\ket$ of $p,\;n,\;p^*$ and $n^*$.  
The experimental values are in the bracket. 
We employed $b_S=0.96,\;b_A=0.7$ [GeV$^2$].}\label{tab:mu_NR}
\begin{tabular}{ccccc}
\vspace*{-5mm}\\
\noalign{\hrule height 0.8pt}
  & $p$ & $n$ & $p^*$ & $n^*$\\
\hline
$Q$ & $1$ & $0$ & $1$ & $0$\\
$\mu(\kappa=0)$ & $1.5 (2.79)$ & $-0.73(-1.91)$ & $0.82$ & $-0.172$ \\
$\mu(\kappa=2)$ & $1.6 (2.79)$ & $-0.76(-1.91)$ & $2.1$ & $-0.61$\\
$\bra r^2 \ket^{1/2}_E$ [fm] & $0.75 (0.86)$ & - & $0.84$ & -\\
$\bra r^2 \ket_E$ [fm$^2$] &- & $-0.074 (-0.12)$ & - & -$0.046$\\
\hline
\end{tabular}
\end{center}
\end{table}
\section{Summary and conclusion}\label{sec:sum}

We discussed the electromagnetic properties of the nucleon and Roper resonance 
in the chiral quark-diquark model.
We showed that the mass difference between the nucleon and Roper 
resonance was reproduced by the difference of the diquark 
correlations between the scalar and axial-vector channels. 
Thus the appearance of the Roper resonance and its excitation energy 
is generated by the violation of the spin-flavor symmetry caused by 
the diquark correlation. In the present study, we employed the 
quark-diquark model in order to realize this scenario, however we should note that 
the violation of the spin-flavor symmetry is not restricted to the quark-diquark model, hence 
the present study can be applied to other approaches.

With the intrinsic form factors of the scalar and axial-vector 
diquarks we reproduced the reasonable values of the charge radii both of the
proton and neutron.
On the other hand, we obtained the smaller values of the magnetic moments of
them than the experimental values due to the 
small mixing angle. 
The deficiencies were mostly observed in the iso-vector magnetic moments.
We expect that these deficiencies are compensated by the inclusion of the
electromagnetic transition vertex between the scalar and axial-vector diquarks
and of the pionic effects, as suggested by Mineo et al~\cite{Mineo:2002bg}.
The pion cloud probably affects both the electric and magnetic form factors, 
then the present picture for the nucleon and Roper resonance should be 
considered as their core part.

We investigated the electric form factors of the nucleon and Roper resonance,
where we employ several values for the axial-vector diquark size from 
$\bra r^2\ket_A^{1/2}$=0.6 $\sim$0.8 [fm].
For the proton components, the charge radii of the $p^*$ and $p$ are almost 
the same size and it is less sensitive to the diquark size.
While the charge radii of $n^*$ is sensitive to the value of $\bra r^2\ket_A^{1/2}$.
When we employ $\bra r^2\ket_A^{1/2}$=0.82 [fm], the charge radii of 
$n^*$ is almost comparable with those of the nucleon both for the proton and 
neutron components.

In conventional picture of the collective excitation of the Roper resonance,
the charge radii of the Roper resonance are larger than those 
of the nucleon both for the proton and neutron component. 
In the quark-diquark picture, the Roper resonance is a 
spin-partner of the nucleon with the different spin component.
In this case the charge radius of the Roper resonance is smaller than 
those in the collective picture.


\end{document}